\documentclass[twocolumn,showpacs,preprintnumbers,amsmath,amssymb]{revtex4}

 \usepackage{graphicx}
 \usepackage{amsmath}
 \usepackage{amsfonts}
 \usepackage{amssymb}
 \usepackage{pstricks}
 \usepackage{psfrag}
 \usepackage{epsfig}
 \usepackage[ansinew]{inputenc}

\begin{document}

\title{Light control of the flow of phototactic microswimmer suspensions }

\author{Xabel Garcia, Salima Rafa\"i and Philippe Peyla}

\affiliation{Laboratory of Interdisciplinary Physics, UMR 5588, Universit\'{e} Joseph Fourier and CNRS, F-38402 Grenoble, France}

\date{Received: \today / Revised version: (date)}

\begin{abstract}

Some micro-algae are sensitive to light intensity gradients. This property is known as phototaxis: the algae swim toward a light source (positive phototaxis). We use this property to control the motion of micro-algae within a Poiseuille flow using light. The combination of flow vorticity and phototaxis results in a concentration of algae around the center of the flow. Intermittent light exposure allows analysis of the dynamics of this phenomenon and its reversibility. With this phenomenon, we hope to pave the way toward new algae concentration techniques (a bottleneck challenge in hydrogen algal production) and toward the improvement of pollutant bio-detector technology.
\end{abstract}


\maketitle

Nature presents a wide and fascinating array of organisms that can propel themselves in a fluid. 
Among them, a lot of micro-organisms \cite{Ginger2008,Jarrell2008} like spermatozoa, bacteria or micro-algae can move with the help of flagella or cilia \cite{turner,Berg2004} 
and are classified as micro-swimmers \cite{Purcell1977}.
Thanks to the 19th century seminal work of Engelmann \cite{engelmann}
and Pfeffer \cite{pfeffer1,pfeffer2} it is well known that some bacteria, like {\it Escherichia Coli}, move toward or away from certain chemicals \cite{Adler}, a phenomenon known as chemotaxis. 
For example, {\it E. Coli} consumes oxygen and swims along oxygen gradients \cite{Dombrowski2004}. Another phenomenon discovered a long time ago on {\it paramecium} 
- a ciliate protozoa - is gravitaxis,
{\it i.e.} sensitivity to gravity which makes the cells 
swim vertically \cite{machemer}, a property also observed on some algae in combination with flow vorticity, together called gyrotaxis \cite{Kessler1985a,durham}. 
The phototaxis property describes the motion of micro-swimmers along gradients of light intensity \cite{sourcebook} and is used by some marine phytoplankton to move vertically in water
columns in the euphotic zone \cite{Kamykowski1988}.

A very active field of research involves understanding how artificial or natural microswimmer suspensions 
can self-organize under the influence of externally controlled fluid-mediated interactions.
For example, some light-powered autonomous micromotors have shown the ability to provoke schooling behavior in water \cite{ibele}.  In the 1980's, John Kessler \cite{Kessler1985}
combined gravity with theflow of a gravitactic alga suspension. The torque exerted by gravity on each alga leads to their self focusing or to their migration toward the walls of the container depending 
on the vertical orientation of the Poiseuille flow. Self focusing or remixing was observed,
depending on the vertical direction of the Poiseuille flow (downward for self-focusing, upward for mixing). Even in nature, in marine eco-system, thin layers of concentrated phytoplankton
are found in the coastal ocean, a few meters below the surface, and contain cell concentrations up to two orders of magnitude above ambient concentrations. This separation phenomenon 
is due to the hydrodynamic flow that causes gyrotactic trapping \cite{durham}.

In our study, we show that by combining light and a Poiseuille flow, we can provoke easily controlled and reversible
self-focusing of an alga suspension of {\it Chlamydomonas Rheinhardtii}.  The advantage of light is that it controls self-focusing/remixing 
regardless of flow direction. Furthermore, by switching the light on and off, we can reverse from self-focusing to remixing.
After analyzing the dynamics of this phenomenon with a periodical
modulation of light exposure, we measured the dynamic response of {\it Chlamydomonas Reinhardtii} to the presence of light. 
Finally, we demonstrated that the migration of a cell through the flow lines can be simply understood and 
qualitatively described and computed with a simple non-linear model.


{\it Chlamydomonas Reinhardtii}  (CR) ~\cite{sourcebook} is a genus of green alga. It is a
bi-flagellated unicellular organism. CR is used as a
model organism in molecular biology, especially for studies of
flagella motion, chloroplast dynamics, biogenesis and genetics. It
is spheroidal in shape with two anterior flagella moving in a 
back-and-forth movement, producing a jerky breast stroke with a mean
speed of $V_0 \sim 50 \mu m/s$ in a water-like medium \cite{Garcia2011}. Since
the cell radius is $R\sim 5\mu m$, Brownian motion is negligible.
CR is photo-tactic thanks to an eyespot \cite{sourcebook}, which is a structure 
made of carotenoid-filled granules in the cell membrane. Under light exposure, the beating of the cis-flagellum (close to the eyespot) 
is inhibited, unlike the beating of the trans-flagellum (opposite the eyespot) which is enhanced: this means the alga
orients itself and swims toward the light source. CC124-strain cells
were obtained from the IBPC lab in Paris \cite{sandrine}.
Synchronous cultures of CR were grown in a Tris-Acetate Phosphate
medium (TAP) using a $14/10~hr$ light/dark cycle at $22^{o} C$.
Cultures were typically grown for two days under fluorescent
lighting before cells were harvested for experiments. 

Microscopy imaging of
chlamydomonas suspensions was carried out with an Olympus inverted
microscope coupled with a fast  camera used at frame rates up to
200 Hz. Square section channels of $1\times 1$mm were made of pdms using soft lithography techniques \cite{whitesides}. 

\begin{figure}
\includegraphics[width=\columnwidth]{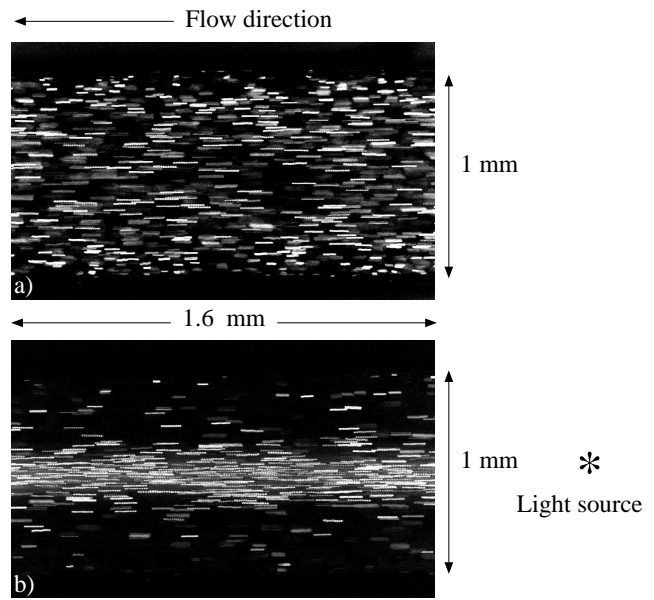}
\caption{\label{trajectories} CR trajectories (obtained by superposing $10$ images). a) With no light. CR are transported by the Poiseuille flow from right to left. 
b) With light on the right hand side: the CR move toward the center.}
\end{figure}

The Reynolds number $Re \sim V_0 R/\nu$ associated with swimming alga in water is very weak (about $2.5\,10^{-4}$ where kinematic viscosity is $\nu \approx 10^{-6}\,m^2s^{-1}$).
It is well known that for vanishing Reynolds numbers, because of Stokes flow reversibility, 
passive spherical particles rotate in a Poiseuille flow (except at the center)
but do not experience any migration across the flow lines \cite{Happel83}. 
In the case of micro-swimmers, the situation is quite different: their motion in a Poiseuille flow
follows an oscillating trajectory \cite{zottl} due to flow vorticity combined with swimming velocity. Therefore, each cell undergoes time periodic cross-stream migration but
with no net migration averaged over the cell's period of rotation. 
Figure~\ref{trajectories}-a shows 40 ms microswimmer tracks in such a Poiseuille flow: the cells are homogeneously distributed over the width of the channel. 
Although the trajectories experience some oscillations, they are not visible at this scale as the spatial wavelength of the oscillations 
is expected to be approximately a few channel widths (see below).

\begin{figure}
\includegraphics[width=\columnwidth]{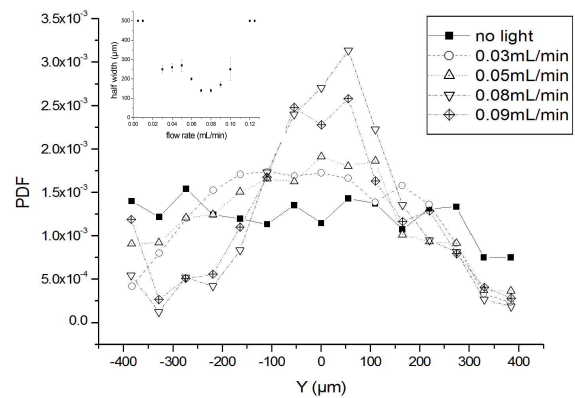}
\caption{\label{PDF} Probability distribution of CR in the Poiseuille Flow. a) without light, b) with light source on the right. 
Inset: The full band width at half maximum as a function of the flow rate.}
\end{figure}

When a light source is switched-on on the right hand side of the flow - {\it i.e.} oriented up-stream since the flow goes from right to left - the situation is noticeably different: the microswimmers migrate to the center of the flow, resulting in strong self-focusing. 
Figure~\ref{trajectories}-b shows 40ms  microswimmer tracks in the presence of light. It represents the final stage where the cells have migrated to the center of the channel. Figures \ref{PDF} shows the corresponding distributions of cells though the channel (without and with light) for different values of the imposed flow rate.
When the light source is on the left hand side - oriented down-stream - with the same flow direction, 
cells migrate toward the walls of the channel.

\begin{figure}
\includegraphics[width=\columnwidth]{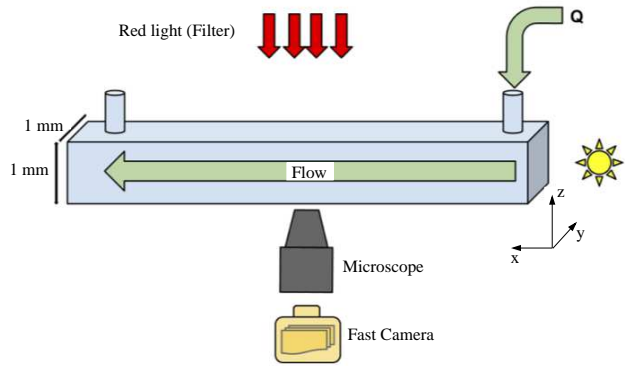}
\caption{\label{setup}The setup of the experiment}
\end{figure}

\begin{figure}
\includegraphics[width=\columnwidth]{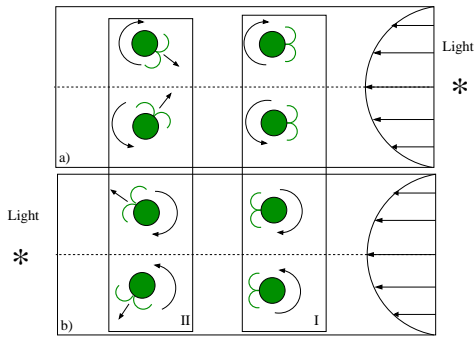}
\caption{\label{poiseuille} Schematic view of the phenomenon. Cells orient themselves toward light (I) and are rotated by the vorticity (II). 
a) Light source upsteram, resulting in a self-focusing. 
b) Light source downstream resulting in a migration toward the wall.}
\end{figure}

Let us now simply sketch the phenomenon. In the presence of light, the swimmer rotates due to flow vorticity, but it also regularly re-orients itself 
toward the light source within a typical time of about 1 second. 
Because of local vorticity, the swimmer will be more frequently oriented by the flow toward the center moving by swimming (fig.\ref{poiseuille}-a). Note that if the light source
is on the opposite side, the cells are more frequently oriented toward the walls to which they migrate (fig.\ref{poiseuille}-b). In the model below, we have solved numerically a simple non-linear model based on \cite{zottl} but including regular swimmer reorientation toward the light. It clearly shows motion toward the center or the walls depending
on the direction of re-orientation {\it i.e.} toward the position of the light source.

Experimentally, the self-focusing is observed in the range of flow rate $0.03 \, ml/min\,<Q<\,0.09\,ml/min$.
Below this range, the flow is too weak to force the cells to rotate since a CR can resist the flow rotation \cite{Rafai2010}, and the algae are not oriented by the flow. Above this range, the cell's rotation is too fast and it has no time to orient itself toward the light, a phenomenon similar to the gyrotactic trap observed in phytoplancton \cite{durham}, and consequently, the cell cannot migrate. The concentration of cells in the center results in a reinforcement of the hydrodynamic interactions between cells, 
this results in a broadening of the distribution of CR around the center where hydrodynamic interactions prevail. As shown on figure (\ref{PDF}), the band width depends on the flow rate. 
The minimum band width obtained at a flow rate of $0.09\,ml/min$ represents $22\%$ of the channel width [see inset of figure(\ref{PDF})].

\begin{figure}
\includegraphics[width=\columnwidth]{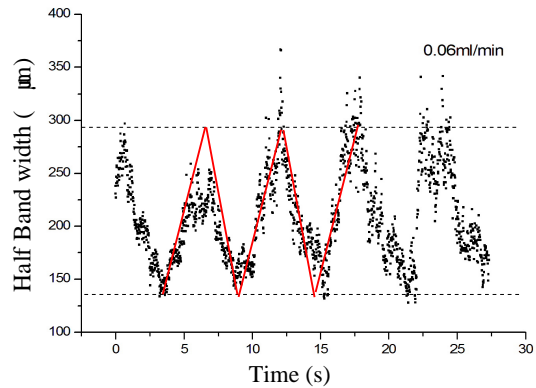}
\caption{\label{dynamics} Band width of CR as a function of time under light exposures of $2.5s$ separated by dark periods of $2.5s$. A linear variation  
is observed with a transverse velocity
equal on average to $\sim60 \mu m s^{-1}$ (indicated by the red lines).}
\end{figure}

Self-focusing is associated with hydrodynamic forcing which orients the cells toward the center of the flow and they migrate at their own velocity. Therefore the phenomenon dynamics are determined by the cell's velocity $V_0$. When the light is switched off from the focusing state, hydrodynamic interactions between oriented CR \cite{evans} reinforced by the concentration of algae
at the flow center, cause
the cells to mix back in the fluid and to fill the channel at the same velocity $V_0$ which dominates over a much slower hydrodynamic diffusion. In figure (\ref{dynamics}), 
we analyze the dynamics of such a phenomenon by varying light exposure time: the light is switched on and off alternately with a time period $T=5s$. Fast camera visualization clearly shows that the half band width varies linearly and reversibly with time. We found that the corresponding average velocity 
for both self-focusing and re-mixing is $60 \mu m s^{-1}$ for a flow rate of $0.06ml/min$. This velocity value is consistent with the usual value of CR.


We then developped a very simple non-linear model that describes the motion of a swimmer within a Poiseuille flow in a cylinder of radius $R$ with regular orientation
toward a light source situated on the right or left hand side of the channel. We do not describe the re-mixing obtained experimentally which is due to a strong repulsion between oriented
CR.
This model is inspired by \cite{zottl} and has been modified in order to
take into account re-orientation of the micro-swimmer upstream or downstream toward a light source.
We define $\bold{e_r}$ and $\bold{k}$ the unitary radial and longitudinal vectors respectively. The micro-swimmer situated in 
$\bold{r}=\rho \, \bold{e_r}+z \, \bold{k}$ is simply seen as moving at velocity $\bold{V}=\bold{V_0}+\bold{u}(\rho)$ 
where $\bold{V_0}$ is the intrinsic velocity of the microswimmer moving in a fluid flowing at velocity $\bold{u}(\rho)=u_{max}(\rho^2-R^2)\bold{k}$ (Poiseuille flow), 
$u_{max}$ being the maximum value of $\bold{u}(\rho)$ at the center of the flow ($\rho=0$).
The direction of $\bold{V_0}$ 
({\it i.e.} the direction of the swimmer) is determined by local flow vorticity $\omega(\rho)=u_{max}\, \rho /R^2$. The trajectory of the micro-swimmer is thus given by:

\begin{equation}
\label{traj}
\bold{r(t)}=\int_0^t \bold{V(\rho)}d \tau,
\end{equation} 
with $\bold{V}=V_0 \sin \theta(\rho) \bold \, {e_r}+[V_0 \cos \theta(\rho)+u(\rho)] \, \bold{k}$ and 

\begin{equation}
\label{theta}
\bold{\theta(\rho)}=\int_0^t \bold{\omega(\rho)}d \tau,
\end{equation}

\begin{figure}
\includegraphics[width=\columnwidth]{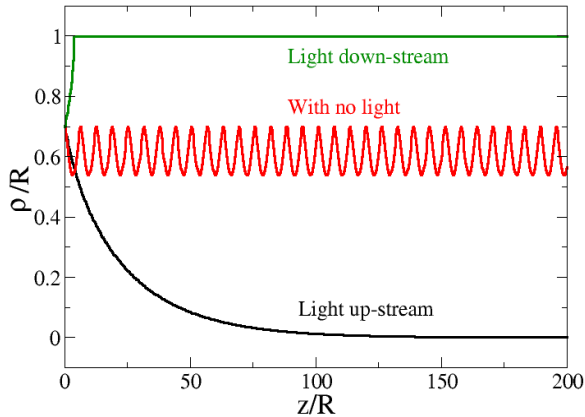}
\caption{\label{traj_th} Trajectories of a micro-swimmer within a Poiseuille flow. $\rho/R=0$ is the center of the Poiseuille flow. 
The micro-swimmers experience oscillations in the absence of orientation toward light (red curve). 
Light can be up or down stream oriented. The micro-swimmers migrate toward the walls if they are regularly oriented down-stream, or they migrate toward the center if oriented regularly up-stream..
}
\end{figure}

By solving this system of equations numerically, we obtain a trajectory which experiences oscillations but no net migration [see figure (\ref{traj_th})]. 
Note that the oscillations have a wavelength much
larger than the radius $R$ \cite{zottl} and are not visible on figure \ref{trajectories}. The oscillations result from the combination of flow vorticity and swimmer velocity. The micro-swimmer is then regurlarly reoriented up-stream with a time period $\tau \approx \omega^{-1}_{max}$ which is the same order of magnitude as in the experiments ($\tau \approx 1\,sec.$ and $\omega_{max}\approx1Hz$). 
The swimmer is still rotated by vorticity, however each time, it is oriented up-stream, vorticity makes the swimmer turn toward the center of the flow where it moves to. Conversely, if the light is on the opposite side, then
the swimmer is oriented down-stream every time $\tau$, and vorticity makes it turn
toward the walls. Note that in the experiments, for the self-focusing effect, due to repulsive hydrodynamic and steric interactions, the cells cannot of course concentrate in $r=0$ 
but are scattered in a band width around the center of the flow (figure \ref{PDF}). Therefore, we believe that only a full description of the hydrodynamics coupled with a suspension of modelized swimmers
could help to predict the precise value of the band width
which depends on the flow rate and on the cell concentration. This will be done in the future.

In this work, we show that the coupling of the phototaxis property of the {\it Chlamydomonas Rheinhardtii} micro-alga with a Poiseuille flow can
lead to reversible self-focusing of micro-algae at the center of the flow. This phenomenon could pave the way toward new algal concentration or separation processes
in pipe flows, a field of particular interest in hydrogen production by algae \cite{greenbaum, moraine} for example. The advantage of self-focusing is to avoid cell accumulation at the walls where adhesion can occur leading to irreversible alteration of the suspension. Another possible application could be found in bio-detectors of river pollutants. These devices use micro-algae with phototaxis properties that are very sensitive to trace of pollutants
such as copper ions or PCP \cite{michels}. The velocity associated with self-focusing being averaged on all the cells, it can help to diagnose the
phototaxis efficiency of the whole suspension of swimming CR.
However, this phenomenon also opens several more fundamental questions that should be addressed in the future by more refined
models. Indeed, the hydrodynamic interaction between cells as well as the fluid disturbance created by the swimmers should be
taken into account in order to better describe the competition between self-focusing and hydrodynamic interactions
resulting in a band of concentrated cells at the center of the channel.  

{\it {Acknowledgments.-}}
This work has been supported by the ANR MICMACSWIM. X.G. thanks G. Kittenbergs for help in data analysis and M.Garcia for help in cell cultures.

  \bibliographystyle{apsrev}
\bibliography{chlamybib}

\end{document}